\documentclass[journal]{IEEEtran}

\usepackage{amssymb}
\usepackage{amsfonts}
\usepackage{amsmath}
\usepackage{pdfpages}
\usepackage{amsthm}
\usepackage{graphics}
\usepackage{float}
\usepackage{stfloats}
\usepackage{graphicx}
\usepackage{subfigure}
\font\euler=eusm10
\def \M{\mbox{\euler M}}
\def \L{\mbox{\euler L}}
\def \N{\mbox{\euler N}}
\def \intd{{\rm d}}
\def \mui{\mbox{\boldmath$\mu$}}
\def \sigmai{\mbox{\boldmath$\sigma$}}

\newtheorem{exm}{Example}

\newtheorem{definition}{Definition}

\ifCLASSINFOpdf

\else

\fi

\begin{document}

\title{Estimating Unknown Time-Varying Parameters in Uncertain Differential Equation}
\author{Guidong~Zhang, Yuhong~Sheng\textsuperscript{*}
\thanks{Y. Sheng is with College of Mathematics and System Science, Xinjiang University, Urumqi 830046, China (e-mail: shengyuhong1@sina.com).}
\thanks{G. Zhang is studying at the College of Mathematics and System Science, Xinjiang University, Urumqi 830046, China (e-mail:
zgd@stu.xju.edu.cn).}
}

\maketitle

\begin{abstract}
Uncertain differential equations have a wide range of applications. How to obtain estimated values of unknown parameters in uncertain differential equations through observations has always been a subject of concern and research, many methods have been developed to estimate unknown parameters. However, these parameters are constants. In this paper, the method of least squares estimation is recast for estimating the unknown time-varying parameters in uncertain differential equations. A set of unknown time-varying parameter estimates will be obtained, and then the unknown time-varying parameters will be obtained by regression fitting using the estimated values. Using this method, the uncertain differential equation of blood alcohol concentration in human body after drinking and the uncertain differential equation of COVID-19 are derived.
\end{abstract}

\IEEEpeerreviewmaketitle

\section{Introduction}
\IEEEPARstart {T}{h}{e} tool that people have always used to deal with random events is probability theory. However, due to the variability and complexity of random events, it takes a lot of manpower, material resources and high-tech means to obtain their distribution. In addition, whether the distribution function is close enough to the real frequency in real life remains to be verified, as well as Liu pointed out evolutions of some undetermined phenomena do not behave like randomness, he also pointed out that when emergency arises, e.g. , war, rumour, flood, and earthquake, we often do not have the historical data to deal with them \cite {Liu1}. Under these situations, it is inappropriate to use probability theory in dealing with some problems. Uncertainty theory based on normality, duality, subadditivity, and product axioms is another axiomatic mathematical system to rationally deal with indeterminacy, which was established by Liu \cite {Liu1} in 2007.
\par
In the framework of probability theory, stochastic differential equations have been widely used in the time evolution modeling of dynamic systems under the influence of random noise. Accordingly, the uncertain differential equation proposed by Liu \cite {Liu2} in 2008 is based on uncertainty theory, it is a kind of differential equation driven by Liu processes. Up to now, uncertain differential equation has been widely applied to finance \cite{Liu3}, optimal control \cite {Zhu}, heat conduction \cite{Yang1}, and population model \cite {Zhang}, etc. For more information on uncertain differential equations, please consult Yao's book \cite {Yao1}. The coefficients of these models sometimes contain unknown parameters, thereby, how to estimate the unknown parameters based on the observations values is a critical problem. For the purpose of solving this problem, Sheng et al. \cite {Sheng1} presented a method of least squares estimation for estimating the unknown parameters. Yao and Liu \cite {Yao2} proposed a method of moment estimation based on difference form of uncertain differential equation, due to the moment estimation for unknown parameters in uncertain differential equations is the solution of a system of equations based on moment conditions, however with some sets of observations, this system of equations has no solution, and the moment estimation is invalid. Following that, Liu \cite {Liu5} proposed a generalized moment estimation method with the idea of solving the optimal value to solve this kind of problem. Lio and Liu \cite {Lio1} rewrote the moment estimation method to estimate the parameters. In addition, Lio et al. \cite {Lio2} proposed the method of uncertain maximum likelihood to estimate the unknown parameters. Later, Yang et al. \cite {Yang2} proposed a method to estimate the unknown parameters of uncertain differential equation from the discretely sampled data via the $\alpha$-path. Sheng and Zhang \cite {Sheng2} introduced three parameter estimation methods based on different forms of solutions. Some achievements have also been made in parameter estimation of some special uncertain differential equations. Zhang and Sheng \cite {Zhang1} used the least square estimation method to estimate the unknown parameters of the uncertain delay differential equation. Zhang et al. \cite {Zhang2} also estimated the parameters of high-order uncertain differential equation. For all of these methods, the unknown parameters are constants. How to estimate the parameters of uncertain differential equations based on observed data when the parameters are time-varying, this is the problem of time-varying parameter estimation.
\par
Estimating the unknown parameters of uncertain differential equations under the observations is a topic that everyone is keen on. On the one hand, because uncertain differential equations have a wide range of applications, it is necessary to obtain its specific form; on the other hand, the parameters themselves can also be deeply studied. Hence, time-varying parameters are proposed, parameters change over time are more consistent with the actual situation in real life, so the estimation of time-varying parameters is useful and is also the content of this paper. In this paper, we find the new method for uncertain differential equation to estimate time-varying parameters. The rest of this paper is organized as follows. In Section 2, we introduce some basic concepts and theorems about uncertain variables and uncertain differential equations. In Section 3, the method of least squares estimation will be recast for estimating the time-varying parameters. In Section 4, we will introduce several commonly used regression functions, give the evaluation criteria, and use the least squares estimation method to get the unknown parameters in the regression function, and finally get the time-varying parameters. In Section 5, we apply the proposed estimation method in two numerical examples. Finally, a concise conclusion is given in Section 6.

\section{Preliminary}
In this section, we introduce some concepts and useful theorems about uncertain variables and uncertain differential equations. The following symbol is used this paper:
$$\bigwedge_{i=1}^{\infty}x_{i}:\mathop{\min}\limits_{1\leq i\leq \infty}x_{i}. $$

\begin{definition} \textup{(Liu \cite{Liu1,Liu4}) Let $\L$ be a $\sigma$-algebra
on a nonempty set $\Gamma.$ A set function $\M:$ $\L\rightarrow [0, 1]$
is called an uncertain measure if it satisfies the following
axioms:\\
{Axiom 1}: (Normality Axiom) $\M\{\Gamma\}=1$ for the
universal set $\Gamma.$\\
{Axiom 2}: (Duality Axiom) $\M\{\Lambda\}+\M\{\Lambda^c\}=1$ for
any event $\Lambda$.\\
{Axiom 3}: (Subadditivity Axiom) For every countable sequence of
events $\Lambda_1, \Lambda_2, \cdots,$ we have
$$\M\left\{\bigcup_{i=1}^{\infty}\Lambda_i\right\}\le \sum_{i=1}^{\infty}\M\left\{\Lambda_i\right\}.$$
{{Axiom 4}: (Product Axiom) Let $(\Gamma_k,\L_k,\M_k)$ be
uncertainty spaces for $k=1, 2, \cdots$, Then the product uncertain
measure $\M$ is an uncertain measure satisfying
$$\M\left\{\prod_{k=1}^{\infty}\Lambda_k\right\}=\bigwedge_{k=0}^{\infty}\M_k\{\Lambda_k\}$$
where $\Lambda_k$ are arbitrarily chosen events from $\L_k$ for
$k=1, 2, \cdots$, respectively.}}
\end{definition}

\begin{definition} \textup{(Liu \cite{Liu1})
An uncertain variable $\xi$ is a measurable function from an uncertain space ($\Gamma$, $\L$, $\M$) to the set of real numbers, i.e., for any Borel set B, the set
$$\{\xi\in B\}=\{\gamma\in\Gamma|\xi(\gamma)\in B\}$$
is an event.}
\end{definition}

\begin{definition} \textup{(Liu \cite{Liu1})
Let $\xi$ be an uncertain variable. Then its uncertainty distribution is defined by
$$\Phi(x)=\M\{\xi\le x\}$$
for any real number $x$.}
\end{definition}

An uncertain variable $\xi$ is called normal if it has an uncertainty distribution $$\Phi(x)=\left(1+\exp\left(\frac{\pi(\mu-x)}{\sqrt3\sigma}\right)\right)^{-1},\quad x\in\Re.$$
Denoted by $\N(\mu,\sigma)$. If $\mu=0$ and $\sigma=1$, then $\xi$ is called a standard normal uncertain variable.
The inverse uncertainty distribution of a standard normal uncertain variable is $$\Phi^{-1}(\alpha)=\frac{\sqrt{3}}{\pi}\ln\frac{\alpha}{1-\alpha},\quad \alpha\in(0, 1).$$

\begin{definition}\textup{(Liu \cite{Liu1})
Let $\xi$ be an uncertain variable, and $k$ be a positive integer. Then the $k$-th moment of $\xi$ is defined by
$$E[\xi^k]=\int_0^{+\infty}\M\left\{\xi^k\ge r\right\}\intd r-\int_{-\infty}^0\M\left\{\xi^k\le r\right\}\intd r$$
provided that at least one of the two integrals is finite.}
\end{definition}
Liu \cite{Liu1} proved that if $\xi$ has an inverse uncertainty distribution $\Phi^{-1}(\alpha)$, then
$$E[\xi^k]=\int_0^1\left(\Phi^{-1}(\alpha)\right)^k\intd\alpha.$$
When $k=1$, this is the expected value. And the variable of $\xi$ is defined by
$$V[\xi]=E[(\xi-E[\xi])^2].$$

An uncertain process is a sequence of uncertain variables indexed by the time. As an uncertain counterpart of Wiener process, Liu process is one of the most frequently used uncertain processes.
\begin{definition}\label{liuprocess}\textup{(Liu \cite{Liu4})
An uncertain process $C_t$ is called a Liu process if\\
(i) $C_0=0$ and almost all sample paths are Lipschitz continuous,\\
(ii) $C_t$ has stationary and independent increments, \\
(iii) the increment $C_{s+t}-C_s$ has a normal uncertainty
distribution  $$\Phi_t(x)= \left(1+\exp\left(-\frac{\pi
x}{\sqrt{3}t}\right)\right)^{-1},\quad x\in\Re.$$}
\end{definition}

Let $X_t$ be an uncertain process. Then the uncertain integral of $X_t$ with respect to the Liu process $C_t$ is
\begin{equation*}
\int_a^bX_t\intd C_t=\lim_{\Delta\to
0}\sum_{i=1}^kX_{t_{i}}\cdot(C_{t_{i+1}}-C_{t_{i}})
\end{equation*}
provided that the limit exists almost surely and is finite for any partition of closed interval $[a, b]$ with
$a=t_1<t_2<\cdots<t_{k+1}=b$ and$$\Delta=\max_{1\le i\le k}|t_{i+1}-t_i|.$$

\begin{definition}\textup{(Liu \cite{Liu4})
Suppose that $C_t$ is a Liu process, and $f$ and $g$ are two measurable real
functions. Then
\begin{equation}\label{udee}
\intd X_t = f(t, X_t)\intd t + g(t, X_t)\intd C_t
\end{equation}
is called an uncertain differential equation.}
\end{definition}

An uncertain process $X_t$ is called the solution of the uncertain differential equation (\ref{udee}) if it satisfies
$$X_t=X_0+\int_0^tf(s, X_s)\intd s+\int_0^tg(s, X_s)\intd C_s. $$
A real-valued function $X_t^{\alpha}$ is called the $\alpha$-path of the uncertain differential equation (\ref{udee}) if it solves the corresponding ordinary differential equation
$$\intd X_t^{\alpha} = f(t,X_t^{\alpha})\intd t + |g(t,X_t^{\alpha})|\Phi^{-1}(\alpha)\intd t$$
where $$\Phi^{-1}(\alpha)=\frac{\sqrt{3}}{\pi}\ln\frac{\alpha}{1-\alpha},\quad \alpha\in(0,1)$$
is the inverse uncertainty distribution of a standard normal uncertain variable.

\begin{definition}\textup{(Chen and Liu \cite{Chen2})
Let $u_{1t}$, $u_{2t}$, $v_{1t}$, $v_{2t}$
be integrable uncertain processes. Then the linear uncertain differential equation
$$\intd X_t=(u_{1t}X_t+u_{2t})\intd t+(v_{1t}X_t+v_{2t}) \intd C_t$$
has a solution
$$X_t=U_t\left(X_0+\int_0^t {\frac{u_{2s}}{U_s}}\intd s+\int_0^t \frac{v_{2s}}{U_s}\intd C_s\right)$$
where
$$U_t=\exp\left(\int_0^t u_{1s} \intd s +\int_0^t v_{1s}\intd C_{s}\right).$$}
\end{definition}

\section{Parameter Estimation}
In this section, we present a new parameter estimation method based on least square estimation to estimate the parameters varying with time in an uncertain differential equation based on  some discrete observations.
\begin{equation}\label{diff}
\intd X_t=f(t, X_t; \mui_{t})\intd t+g(t, X_t; \sigmai_{t})\intd C_t
\end{equation}
where $\mui_{t}$ and $\sigmai_{t}$ are unknown time-varying parameters to be estimated. We use a method, the means of least squares estimation will be recast as follows.

First, we have $N$ observations $x_{t_{i}}$ $(i=1, 2, \cdots, N)$, let us estimate $\mui_{t_{m}}$, $\sigmai_{t_{m}}$, $(m=1, 2, \cdots, N-n+1)$ by applying $n$ observed data $x_{t_{m}}$, $x_{t_{m+1}}$, $\cdots$, $x_{t_{m+n-1}}$. A frequently used numerical approximation to the equation (\ref{diff}) is the Euler approximate
$$X_{t_{i+1}}=X_{t_i}+f(t_i, X_{t_i}; \mui_{t_{m}})\cdot{(t_{i+1}-t_i)}$$
\begin{equation}\label{euler}
+g(t_i, X_{t_i}; \sigmai_{t_{m}})\cdot{(C_{t_{i+1}}-C_{t_i})}
\end{equation}
which could be equivalent to the following form
$$X_{t_{i+1}}-X_{t_i}-f(t_i, X_{t_i}; \mui_{t_{m}})\cdot{(t_{i+1}-t_i)}$$
\begin{equation}\label{noise}
=g(t_i, X_{t_i}; \sigmai_{t_{m}})\cdot{(C_{t_{i+1}}-C_{t_i})}.
\end{equation}
According to the method of least squares estimation, note that the right term in the equation (\ref{noise}) is usually regarded as the noise, which should be as small as possible. Hence, give the observed data ($t_{i}$, $x_{t_{i}}$), $(i=m, m+1, \cdots, m+n-1)$, the estimate of $\mui_{t_{m}}$ solves the following optimization problem
\begin{equation}\label{opt}
\mathop{\min}\limits_{\mui_{t_{m}}} \sum_{i=m}^{m+n-2}(x_{t_{i+1}}-x_{t_i}-f(t_i, x_{t_i}; \mui_{t_{m}})\cdot{(t_{i+1}-t_i)})^2. \end{equation}
Let $\tilde{\mui}_{t_{m}}$ denote the estimate of $\mui_{t_{m}}$ obtained from the optimization problem (\ref{opt}). Then, the estimate of $\sigmai_{t_{m}}$ solves the following equation:
$$
E\left[\sum_{i=m}^{m+n-2}\left(g(t_i, x_{t_i}; \sigmai_{t_{m}})\cdot(C_{t_{i+1}}-C_{t_i})\right)^2\right]$$
$$=\sum_{i=m}^{m+n-2}\left(x_{t_{i+1}}-x_{t_i}-f(t_i, x_{t_i}; \tilde{\mui}_{t_{m}})\cdot(t_{i+1}-t_i)\right)^2
$$
where $C_{t}$ is a Liu process, and the increment $C_{t_{i+1}}-C_{t_i}$ is a normal uncertain variable with an expected value 0 and variance $(t_{i+1}-t_{i})^2$, we have
\begin{align*}
&E\left[\sum_{i=m}^{m+n-2}\left(g(t_i, x_{t_i}; \sigmai_{t_{m}})\cdot(C_{t_{i+1}}-C_{t_i})\right)^2\right]\\[2mm]
=&\sum_{i=m}^{m+n-2}E\left[\left(g(t_i,x_{t_i};\sigmai_{t_{m}})\cdot(C_{t_{i+1}}-C_{t_i})\right)^2\right]\\[2mm]
=&\sum_{i=m}^{m+n-2}g(t_i, x_{t_i}; \sigmai_{t_{m}})^2\cdot E\left[(C_{t_{i+1}}-C_{t_i})^2\right]\\[2mm]
=&\sum_{i=m}^{m+n-2}g(t_i, x_{t_i}; \sigmai_{t_{m}})^2\cdot\left(t_{i+1}-t_i\right)^2
\end{align*}
therefore, the estimate of $\sigmai_{t_{m}}$ is a solution of the following equation
$$ \sum_{i=m}^{m+n-2}g(t_i, x_{t_i}; \sigmai_{t_{m}})^2\cdot{(t_{i+1}-t_{i})^2}$$
\begin{equation}\label{si}
=\sum_{i=m}^{m+n-2}(x_{t_{i+1}}-x_{t_i}-f(t_i, x_{t_i}; \tilde{\mui}_{t_{m}})\cdot{(t_{i+1}-t_i)})^2.
\end{equation}
We have the estimate $(\tilde{\mui}_{t_{m}}, \tilde{\sigmai}_{t_{m}})$, as an analogy, we can get the estimate values $(\tilde{\mui}_{t_{m+1}}, \tilde{\sigmai}_{t_{m+1}})$, $(\tilde{\mui}_{t_{m+2}},\tilde{\sigmai}_{t_{m+2}})$, $\cdots$,\\ $(\tilde{\mui}_{t_{N-n+1}}, \tilde{\sigmai}_{t_{N-n+1}})$, where $\tilde{\mui}_{t_{m}}=(\tilde{\mu}_{1t_{m}}, \tilde{\mu}_{2t_{m}}, \cdots, \tilde{\mu}_{j_{1}t_{m}})$ and $\tilde{\sigmai}_{t_{m}}=(\tilde{\sigma}_{1t_{m}}, \tilde{\sigma}_{2t_{m}}, \cdots, \tilde{\sigma}_{j_{2}t_{m}})$.
Through the above analysis, we get a set of observations by means of least squares estimation, the following we just need to fit this set of observations to get the specific form of $\mui_{t}$ and $\sigmai_{t}$.

\begin{exm}\label{ex1}\textup{Consider the uncertain differential equation
$$\intd X_t=\mu_{t}X_{t}\intd t+\sigma_{t}X_{t}\intd C_t$$
the estimate $\tilde{\mu}_{t{m}}$ solves the optimization problem
$$\min_{\mu_{t_{m}}}\sum_{i=m}^{m+n-2}\left(x_{t_{i+1}}-x_{t_i}-\mu_{t_{m}}\cdot x_{t_i}(t_{i+1}-t_i)\right)^2$$
that is
$$\tilde{\mu}_{t_{m}}=\left(\sum_{i=m}^{m+n-2}(x_{t_{i+1}}-x_{t_{i}})(t_{i+1}-t_{i})x_{t_{i}}\right)\times\\$$ $$\left(\sum_{i=m}^{m+n-2}x_{t_{i}}^2(t_{i+1}-t_{i})^2\right)^{-1}$$
the estimate $\tilde{\mu}_{t_{m+1}}$ solves the optimization problem
$$\mathop{\min}\limits_{\mu_{t_{m+1}}} \sum_{i=m+1}^{m+n-1}(x_{t_{i+1}}-x_{t_i}-\mu_{t_{m+1}}\cdot x_{t_{i}}{(t_{i+1}-t_i)})^2$$
that is
$$\tilde{\mu}_{t_{m+1}}=\left(\sum_{i=m+1}^{m+n-1}(x_{t_{i+1}}-x_{t_{i}})(t_{i+1}-t_{i})x_{t_{i}}\right)\times\\$$ $$\left(\sum_{i=m+1}^{m+n-1}x_{t_{i}}^2(t_{i+1}-t_{i})^2\right)^{-1}$$
we have the estimate values $\tilde{\mu}_{t_{m+2}}$, $\tilde{\mu}_{t_{m+3}}$, $\cdots$, $\tilde{\mu}_{t_{N-n+1}}$.
Then, the parameter $\tilde{\sigma}_{t{m}}$ satisfies
$$\sum_{i=m}^{m+n-2}\sigma_{t_{m}}^2\cdot x_{t_{i}}^2{(t_{i+1}-t_{i})^2}$$$$=\sum_{i=m}^{m+n-2}(x_{t_{i+1}}-x_{t_i}-\tilde{\mu}_{t_{m}}\cdot x_{t_{i}}{(t_{i+1}-t_i)})^2$$
this is equivalent to
$$\tilde{\sigma}_{t_{m}}=\left(\frac{\sum_{i=m}^{m+n-2}(x_{t_{i+1}}-x_{t_{i}})^2}{\sum_{i=m}^{m+n-2}x_{t_{i}}^2(t_{i+1}-t_{i})^2}-\tilde{\mu}_{t_{m}}^2\right)^{\frac{1}{2}}$$
the same
$$\tilde{\sigma}_{t_{m+1}}=\left(\frac{\sum_{i=m+1}^{m+n-1}(x_{t_{i+1}}-x_{t_{i}})^2}{\sum_{i=m+1}^{m+n-1}x_{t_{i}}^2(t_{i+1}-t_{i})^2}-\tilde{\mu}_{t_{m+1}}^2\right)^{\frac{1}{2}}$$
we have the estimate values $\tilde{\sigma}_{t_{m+2}}$, $\tilde{\sigma}_{t_{m+3}}$, $\cdots$, $\tilde{\sigma}_{t_{N-n+1}}$.}
\end{exm}

\begin{exm}\label{ex2}\textup{Consider the uncertain differential equation
$$\intd X_t=(\mu_{1t}+\mu_{2t}X_{t})\intd t+\sigma_{t}\intd C_t$$
where $\mu_{1t}$, $\mu_{2t}$, and $\sigma_{t}>0$ are parameters  to be estimated. According to (\ref{opt}), the estimates $\tilde{\mu}_{1t_{m}}$ and $\tilde{\mu}_{2t_{m}}$ solve the optimization problem
$$\mathop{\rm{min}}_{
        \mu_{1t_{m}}
        \atop
        \mu_{2t_{m}}} \sum_{i=m}^{m+n-2}(x_{t_{i+1}}-x_{t_i}-(\mu_{1t_{m}}+\mu_{2t_{m}}x_{t_i})\cdot{(t_{i+1}-t_i)})^2$$
that is
$\left[\begin{array}{c}
\tilde{\mu}_{1t_{m}}\\[3mm]
\tilde{\mu}_{2t_{m}}
\end{array}\right]=$$$
\left[\begin{array}{cc}
\sum\limits_{i=m}^{m+n-2}(t_{i+1}-t_i)^2 &\sum\limits_{i=m}^{m+n-2}x_{t_i}(t_{i+1}-t_i)^2\\[3mm]
\sum\limits_{i=m}^{m+n-2}x_{t_i}(t_{i+1}-t_i)^2 &\sum\limits_{i=m}^{m+n-2}x_{t_i}^2(t_{i+1}-t_i)^2
\end{array}\right]^{-1}
$$
$$\times
\left[\begin{array}{c}
\sum\limits_{i=m}^{m+n-2}(x_{t_{i+1}}-x_{t_i})(t_{i+1}-t_i)\\[3mm]
\sum\limits_{i=m}^{m+n-2}x_{t_i}(x_{t_{i+1}}-x_{t_i})(t_{i+1}-t_i)
\end{array}\right]
$$
the estimates $\tilde{\mu}_{1t_{m+1}}$ and $\tilde{\mu}_{2t_{m+1}}$ solve the optimization problem
$$\mathop{\rm{min}}_{
        \mu_{1t_{m+1}}
        \atop
        \mu_{2t_{m+1}}}\sum_{i=m+1}^{m+n-1}(x_{t_{i+1}}-x_{t_i}-(\mu_{1t_{m+1}}+\mu_{2t_{m+1}}x_{t_i})\cdot{(t_{i+1}-t_i)})^2$$
that is
$\left[\begin{array}{c}
\tilde{\mu}_{1t_{m+1}}\\[3mm]
\tilde{\mu}_{2t_{m+1}}
\end{array}\right]
=$$$
\left[\begin{array}{cc}
\sum\limits_{i=m+1}^{m+n-1}(t_{i+1}-t_i)^2 &\sum\limits_{i=m+1}^{m+n-1}x_{t_i}(t_{i+1}-t_i)^2\\[3mm]
\sum\limits_{i=m+1}^{m+n-1}x_{t_i}(t_{i+1}-t_i)^2 &\sum\limits_{i=m+1}^{m+n-1}x_{t_i}^2(t_{i+1}-t_i)^2
\end{array}\right]^{-1}
$$
$$\times
\left[\begin{array}{c}
\sum\limits_{i=m+1}^{m+n-1}(x_{t_{i+1}}-x_{t_i})(t_{i+1}-t_i)\\[3mm]
\sum\limits_{i=m+1}^{m+n-1}x_{t_i}(x_{t_{i+1}}-x_{t_i})(t_{i+1}-t_i)
\end{array}\right]
$$
we have the estimate values $\tilde{\mu}_{1t_{m+2}}$, $\tilde{\mu}_{2t_{m+2}}$, $\tilde{\mu}_{1t_{m+3}}$, $\tilde{\mu}_{2t_{m+3}}$, $\cdots$, $\tilde{\mu}_{1t_{N-n+1}}$, $\tilde{\mu}_{2t_{N-n+1}}$.
Then, according to (\ref{si}), the parameters $\tilde{\sigma}_{t_{m}}$ satisfies
$$ \sum_{i=m}^{m+n-2}\sigma_{t_{m}}^2\cdot{(t_{i+1}-t_{i})^2}$$$$=\sum_{i=m}^{m+n-2}(x_{t_{i+1}}-x_{t_i}-(\tilde{\mu}_{1t_{m}}+\tilde{\mu}_{2t_{m}}x_{t_{i}})\cdot{(t_{i+1}-t_i)})^2$$
that is \\
$\tilde{\sigma}_{t_{m}}$$$=\left(\frac{\sum\limits_{i=m}^{m+n-2}(x_{t_{i+1}}-x_{t_i}-(\tilde{\mu}_{1t_{m}}+\tilde{\mu}_{2t_{m}}x_{t_{i}})\cdot{(t_{i+1}-t_i)})^2}{\sum\limits_{i=m}^{m+n-2}{(t_{i+1}-t_{i})^2}}\right)^{\frac{1}{2}}$$
the parameters $\tilde{\sigma}_{t_{m+1}}$ satisfies
$$ \sum_{i=m+1}^{m+n-1}\sigma_{t_{m+1}}^2\cdot{(t_{i+1}-t_{i})^2}$$$$=\sum_{i=m+1}^{m+n-1}(x_{t_{i+1}}-x_{t_i}-(\tilde{\mu}_{1t_{m+1}}+\tilde{\mu}_{2t_{m+1}}x_{t_{i}})\cdot{(t_{i+1}-t_i)})^2$$
that is \\
$\tilde{\sigma}_{t_{m+1}}=$$$\left(\frac{\sum\limits_{i=m+1}^{m+n-1}(x_{t_{i+1}}-x_{t_i}-(\tilde{\mu}_{1t_{m+1}}+\tilde{\mu}_{2t_{m+1}}x_{t_{i}})\cdot{(t_{i+1}-t_i)})^2}{\sum\limits_{i=m+1}^{m+n-1}{(t_{i+1}-t_{i})^2}}\right)^{\frac{1}{2}}$$
we have the estimate values $\tilde{\sigma}_{1t_{m+2}}$, $\tilde{\sigma}_{2t_{m+2}}$, $\tilde{\sigma}_{1t_{m+3}}$, $\tilde{\sigma}_{2t_{m+3}}$, $\cdots$, $\tilde{\sigma}_{1t_{N-n+1}}$, $\tilde{\sigma}_{2t_{N-n+1}}$.}
\end{exm}

\begin{exm}\label{ex3}\textup{Consider the uncertain differential equation
$$\intd X_t=(\mu_{1t}+\mu_{2t}X_{t})\intd t+(\sigma_{1t}+\sigma_{2t})X_{t}\intd C_t$$
where $\mu_{1t}$, $\mu_{2t}$, $\sigma_{1t}>0$ and $\sigma_{2t}>0$ are parameters  to be estimated. According to (\ref{opt}), the estimates $\tilde{\mu}_{1t_{m}}$ and $\tilde{\mu}_{2t_{m}}$ solve the optimization problem
$$\mathop{\rm{min}}_{
        \mu_{1t_{m}}
        \atop
        \mu_{2t_{m}}} \sum_{i=m}^{m+n-2}(x_{t_{i+1}}-x_{t_i}-(\mu_{1t_{m}}+\mu_{2t_{m}}x_{t_i})\cdot{(t_{i+1}-t_i)})^2$$
that is
$\left[\begin{array}{c}
\tilde{\mu}_{1t_{m}}\\[3mm]
\tilde{\mu}_{2t_{m}}
\end{array}\right]=$$$
\left[\begin{array}{cc}
\sum\limits_{i=m}^{m+n-2}(t_{i+1}-t_i)^2 &\sum\limits_{i=m}^{m+n-2}x_{t_i}(t_{i+1}-t_i)^2\\[3mm]
\sum\limits_{i=m}^{m+n-2}x_{t_i}(t_{i+1}-t_i)^2 &\sum\limits_{i=m}^{m+n-2}x_{t_i}^2(t_{i+1}-t_i)^2
\end{array}\right]^{-1}
$$
$$\times
\left[\begin{array}{c}
\sum\limits_{i=m}^{m+n-2}(x_{t_{i+1}}-x_{t_i})(t_{i+1}-t_i)\\[3mm]
\sum\limits_{i=m}^{m+n-2}x_{t_i}(x_{t_{i+1}}-x_{t_i})(t_{i+1}-t_i)
\end{array}\right]
$$
the estimates $\tilde{\mu}_{1t_{m+1}}$ and $\tilde{\mu}_{2t_{m+1}}$ solve the optimization problem
$$\mathop{\rm{min}}_{
        \mu_{1t_{m+1}}
        \atop
        \mu_{2t_{m+1}}}\sum_{i=m+1}^{m+n-1}(x_{t_{i+1}}-x_{t_i}-(\mu_{1t_{m+1}}+\mu_{2t_{m+1}}x_{t_i})\cdot{(t_{i+1}-t_i)})^2$$
that is
$\left[\begin{array}{c}
\tilde{\mu}_{1t_{m+1}}\\[3mm]
\tilde{\mu}_{2t_{m+1}}
\end{array}\right]
=$$$
\left[\begin{array}{cc}
\sum\limits_{i=m+1}^{m+n-1}(t_{i+1}-t_i)^2 &\sum\limits_{i=m+1}^{m+n-1}x_{t_i}(t_{i+1}-t_i)^2\\[3mm]
\sum\limits_{i=m+1}^{m+n-1}x_{t_i}(t_{i+1}-t_i)^2 &\sum\limits_{i=m+1}^{m+n-1}x_{t_i}^2(t_{i+1}-t_i)^2
\end{array}\right]^{-1}
$$
$$\times
\left[\begin{array}{c}
\sum\limits_{i=m+1}^{m+n-1}(x_{t_{i+1}}-x_{t_i})(t_{i+1}-t_i)\\[3mm]
\sum\limits_{i=m+1}^{m+n-1}x_{t_i}(x_{t_{i+1}}-x_{t_i})(t_{i+1}-t_i)
\end{array}\right]
$$
we have the estimate values $\tilde{\mu}_{1t_{m+2}}$, $\tilde{\mu}_{2t_{m+2}}$, $\tilde{\mu}_{1t_{m+3}}$, $\tilde{\mu}_{2t_{m+3}}$, $\cdots$, $\tilde{\mu}_{1t_{N-n+1}}$, $\tilde{\mu}_{2t_{N-n+1}}$.
Then, according to (\ref{si}), the parameters $\tilde{\sigma}_{1t_{m}}$ and $\tilde{\sigma}_{2t_{m}}$ satisfy
$$ \sum_{i=m}^{m+n-2}((\sigma_{1t_{m}}+\sigma_{2t_{m}})x_{t_{i}})^2\cdot{(t_{i+1}-t_{i})^2}$$$$=\sum_{i=m}^{m+n-2}(x_{t_{i+1}}-x_{t_i}-(\tilde{\mu}_{1t_{m}}+\tilde{\mu}_{2t_{m}}x_{t_{i}})\cdot{(t_{i+1}-t_i)})^2$$
this is equivalent to
$$(\tilde{\sigma}_{1t_{m}}+\tilde{\sigma}_{2t_{m}})^2=M_{1}$$
where
$$M_{1}=\frac{\sum\limits_{i=m}^{m+n-2}(x_{t_{i+1}}-x_{t_i}-(\tilde{\mu}_{1t_{m}}+\tilde{\mu}_{2t_{m}}x_{t_{i}})\cdot{(t_{i+1}-t_i)})^2}{\sum\limits_{i=m}^{m+n-2}{(x_{t_{i}}(t_{i+1}-t_{i}))^2}}$$
in order to get $\tilde{\sigma}_{1t_{m}}$ and $\tilde{\sigma}_{2t_{m}}$, if there exist weights $\omega_{11}$ and $\omega_{12}$, $(\omega_{11}, \omega_{12}\in[0,1])$ we have
$$\tilde{\sigma}_{1t_{m}}=\omega_{11}M_{1}^{\frac{1}{2}}, \quad \tilde{\sigma}_{2t_{m}}=\omega_{12}M_{1}^{\frac{1}{2}}$$
where $\omega_{11}+ \omega_{12}=1$.
The parameters $\tilde{\sigma}_{1t_{m+1}}$ and $\tilde{\sigma}_{2t_{m+1}}$ satisfy
$$ \sum_{i=m+1}^{m+n-1}((\sigma_{1t_{m+1}}+\sigma_{2t_{m+1}})x_{t_{i}})^2\cdot{(t_{i+1}-t_{i})^2}$$$$=\sum_{i=m+1}^{m+n-1}(x_{t_{i+1}}-x_{t_i}-(\tilde{\mu}_{1t_{m+1}}+\tilde{\mu}_{2t_{m+1}}x_{t_{i}})\cdot{(t_{i+1}-t_i)})^2$$
this is equivalent to
$$(\tilde{\sigma}_{1t_{m+1}}+\tilde{\sigma}_{2t_{m+1}})^2=M_{2}$$
where\\
$M_{2}=$$$\frac{\sum\limits_{i=m+1}^{m+n-1}(x_{t_{i+1}}-x_{t_i}-(\tilde{\mu}_{1t_{m+1}}+\tilde{\mu}_{2t_{m+1}}x_{t_{i}})\cdot{(t_{i+1}-t_i)})^2}{\sum\limits_{i=m+1}^{m+n-1}{(x_{t_{i}}(t_{i+1}-t_{i}))^2}}$$
we have
$$\tilde{\sigma}_{1t_{m+1}}=\omega_{21}M_{2}^{\frac{1}{2}}, \quad \tilde{\sigma}_{2t_{m+1}}=\omega_{22}M_{2}^{\frac{1}{2}}$$
where $\omega_{21}+ \omega_{22}=1$. We have the estimate values $\tilde{\sigma}_{1t_{m+2}}$, $\tilde{\sigma}_{2t_{m+2}}$, $\tilde{\sigma}_{1t_{m+3}}$, $\tilde{\sigma}_{2t_{m+3}}$, $\cdots$, $\tilde{\sigma}_{1t_{N-n+1}}$, $\tilde{\sigma}_{2t_{N-n+1}}$. At this point, the estimates of the two time-varying parameters at different times are estimated.}
\end{exm}

\section{Regression Analysis}
We know that the unknown time-varying parameters $\mui_{t}$ and $\sigmai_{t}$ are functions of time, and a set of estimated values of $\mui_{t}=(\mu_{1t}, \mu_{2t}, \cdots, \mu_{j_{1}t})$ and $\sigmai_{t}=(\sigma_{1t}, \sigma_{2t}, \cdots, \sigma_{j_{2}t})$ can be obtained through the above process. In this part, linear fitting and nonlinear fitting are introduced, several common nonlinear regression models will be presented.\\
(\uppercase\expandafter{\romannumeral1}) Linear regression equation: $$\mu_{jt}=\hat{\beta}_{j0}+\hat{\beta}_{j1}t.  \qquad (j=1, 2, \cdots, j_1)$$
(\uppercase\expandafter{\romannumeral2}) Exponential regression function: $$\mu_{jt}=\hat{\beta}_{j0}\exp(-\hat{\beta}_{j1}t).$$
(\uppercase\expandafter{\romannumeral3}) Logical growth curve function:$$\mu_{jt}=K/(1+\hat{\beta}_{j0}\exp(-\hat{\beta}_{j1}t)).$$

First, we make the scatter diagram, and then choose the appropriate regression function to fit according to the scatter diagram, to evaluate the significance of these models, we consider the coefficient of determination $R^2_j$ with data $(t_{m}, \tilde{\mui}_{t_{m}})$ $m=1, 2, \cdots, N-n+1$, the total sum of squares is $SST_j=\sum_{m=1}^{N-n+1}(\tilde{\mu}_{jt_{m}}-\bar{\mu}_{j})^2$ where $\bar{\mu}_{j}=\frac{1}{N-n+1}\sum_{m=1}^{N-n+1}\tilde{\mu}_{jt_{m}}$, the return to the sum of squares is $SSR_j=\sum_{m=1}^{N-n+1}(\hat{\mu}_{jt_{m}}-\bar{\mu}_{j})^2$ where $\hat{\mu}_{jt_{m}}$ is the fitted value, and the coefficient of determination $R^2_j$ is $R^2_j=\frac{SSR_j}{SST_j}$. The larger the $R^2_j$, the better the curve fitting, where in the best case $R^2_j=1$.

For the above parameters $\hat{\beta}_{ji}$ $(i=0, 1)$, we also use the least squares estimate
$\mathop{\min}\limits_{\hat{\beta}_{ji}}\sum_{m=1}^{N-n+1}(\tilde{\mu}_{jt_{m}}-\hat{\mu}_{jt_{m}})^2 $.

If we use the linear regression function to fit, then we get $\hat{\beta}_{j0}$ and $\hat{\beta}_{j1}$ as follows
\[
\begin{cases}
\hat{\beta}_{j1}= \frac{(N-n+1)\sum_{m=1}^{N-n+1}t_{m}\tilde{\mu}_{jt_{m}}-\sum_{m=1}^{N-n+1}t_{m}\sum_{m=1}^{N-n+1}\tilde{\mu}_{jt_{m}}}{(N-n+1)\sum_{m=1}^{N-n+1}t_{i}^2-(\sum_{m=1}^{N-n+1}t_{i})^2}\\\\
\hat{\beta}_{j0}= \bar{\mu}_j-\hat{\beta}_{j1}\bar{t}
\end{cases} \]
where $\bar{t}=\frac{1}{N-n+1}\sum_{m=1}^{N-n+1}t_{m}$, then, we can get
$$\mu_{jt}=\hat{\beta}_{j0}+\hat{\beta}_{j1}t.$$

If we use nonlinear regression model to fit, Gauss-Newton algorithm is needed, and the algorithm steps are as follows:

\textbf{Step 1.} According to the fitting object, a more appropriate nonlinear fitting function $\mu_j(\hat{\beta}_{ji}, t)$, $(i=0, 1)$ is selected.

\textbf{Step 2.} The partial derivative of the fitting function $\mu_j(\hat{\beta}_{ji}, t)$ is $\frac{\partial \mu_j(\hat{\beta}_{ji}, t)}{\partial \hat{\beta}_{ji}}$.

\textbf{Step 3.} The initial value $\hat{\beta}_{ji}(0)$ of the fitting coefficient is given.

\textbf{Step 4.} Calculate the matrix A and vector B, where
\[A=
\begin{bmatrix}
a_{j00} & a_{j01} \\ a_{j10} & a_{j11}
\end{bmatrix}
\]
$$B=(b_{j0}, b_{j1})^\mathsf{T}$$
where
$$a_{jik}=\sum(\frac{\partial \mu_{j0}(\hat{\beta}_{ji},t_{m})}{\partial \hat{\beta}_{ji}})(\frac{\partial \mu_{j0}(\hat{\beta}_{jk},t_{m})}{\partial \hat{\beta}_{jk}}) \quad   (k=0, 1) $$
$$b_{ji}=\sum(\frac{\partial \mu_{j0}(\hat{\beta}_{ji}, t_{m})}{\partial \hat{\beta}_{ji}})(\tilde{\mu}_{jt_{m}}-\mu_{j0}(\hat{\beta}_{ji}, t_{m}))$$
$$\mu_{j0}(\hat{\beta}_{ji}, t_{m})=\mu_j(\hat{\beta}_{j0}(0), \hat{\beta}_{j1}(0), t_{m}).$$

\textbf{Step 5.} Solve $\delta\hat{\beta}_{ji}$ according to the normal equation $A\times C=B$, where
$$C=(\delta\hat{\beta}_{j0}, \delta\hat{\beta}_{j1})^\mathsf{T}.$$

\textbf{Step 6.} Determining whether $|\delta\hat{\beta}_{ji}|$ is less than the predetermined decimal $\varepsilon$. If condition $\max|\delta\hat{\beta}_{ji}|<\varepsilon$ is true, the iterative calculation will end. Otherwise, \textbf{Step 7} will be carried out.

\textbf{Step 7.} Assign $\hat{\beta}_{ji}(0)+\delta\hat{\beta}_{ji}$ to $\hat{\beta}_{ji}(0)$, return to \textbf{Step 4}.
Through the above algorithm, $\hat{\beta}_{j0}$ and $\hat{\beta}_{j1}$ can be obtained, and $\mu_{jt}$ can be obtained.

\section{Numerical Experiments}
In this section, we will use observed data to illustrate the method of least square estimation in uncertain differential equations with unknown time-varying parameters.

\begin{exm}\label{ex3}\textup{Consider the example of a person's blood alcohol concentration when drinking alcohol. First we derive the uncertain differential equation, suppose a person drinks alcohol with $\mu_{t}$ amount of alcohol in their stomach, alcohol is absorbed by the blood at a rate of $k_{0}$ and consumed at a rate of $k_{1}$, then the blood alcohol concentration with an uncertain disturbance term can be expressed as,
$$X_{t_{i+1}}-X_{t_i}=(k_{0}\mu_{t}-k_{1}X_{t_{i}})(t_{i+1}-t_{i})+\sigma_{t}(C_{t_{i+1}}-C_{t_{i}})$$
generally, during a time interval $[0, t]$ with a partition $0=t_{0}<t_{1}<\cdots<t_{n}=t$, we have
$$X_{t}-X_{0}=\sum_{i=0}^{n-1}(X_{t_{i+1}}-X_{t_{i}})=\sum_{i=0}^{n-1}(k_{0}\mu_{t_{i}}-k_{1}X_{t_{i}})(t_{i+1}-t_{i})$$$$+\sum_{i=0}^{n-1}\sigma_{t_{i}}(C_{t_{i+1}}-C_{t_{i}})$$
with
$$\mathop{\max}\limits_{1\leq i \leq n-1}\mid{t_{i+1}-t_{i}}\mid\rightarrow 0.$$
That is,
$$X_{t}-X_{0}=\int_0^t\mathrm{ (k_{0}\mu_{s}-k_{1}X_{s})}\mathrm{d} s+\int_0^t\mathrm{ \sigma_{s}}\mathrm { d } C_s.$$
Thus we obtain a model of human blood alcohol concentration based on an uncertain differential equation
$$\intd X_{t}=(k_{0}\mu_{t}-k_{1}X_{t})\intd t+\sigma_{t}\intd C_{t}$$
where $\mu_{t}$ and $\sigma_{t}$ are parameters to be estimated. And just to keep things simple, we assume $k_{0}=0.7$, $k_{1}=0.2$. Assume that we have $N=30$ groups of observed data shown in Table \ref{T1}. Let $t_{1}$, $t_{2}$, $\cdots$, $t_{30}$ represent the time for people to observe the blood alcohol concentration after drinking, for example, $t_{2}$ represents observation after 0.25 hours after drinking, let $x_{0}$, $x_{0.25}$, $\cdots$, $x_{16}$ represent the blood alcohol concentration. First of all, let is set $m=i=1$, then $t_1=0$, so
$$X_{t_{i+1}}-X_{t_i}-f(t_i, X_{t_i}; \mu_{0})\cdot{(t_{i+1}-t_i)}$$
$$=g(t_i, X_{t_i}; \sigma_{0})\cdot{(C_{t_{i+1}}-C_{t_i})}.$$
}
\begin{table}
\caption{Observed data in Example \ref{ex3}}\label{T1}
\center
\begin{tabular}{|c|c|c|c|c|c|}
\hline
$i$ &1&2&3&4&5\\
\hline
$t_i$ & 0&0.25&0.3&0.35&0.4\\
\hline
$x_{t_i}$& 0&30&39&46&52\\
\hline
$i$ &6&7&8&9&10\\
\hline
$t_i$&0.45&0.5&0.55&0.6&0.7\\
\hline
$x_{t_i}$&60&68&70&72&74\\
\hline
$i$ &11&12&13&14&15\\
\hline
$t_i$ & 0.75&1&1.5&2&3\\
\hline
$x_{t_i}$&75&80&80&77&68\\
\hline
$i$ &16&17&18&19&20\\
\hline
$t_i$ & 3.5&4&4.5&5&6\\
\hline
$x_{t_i}$&58&51&50&45&38\\
\hline
$i$ &21&22&23&24&25\\
\hline
$t_i$ & 7&8&9&10&11\\
\hline
$x_{t_i}$&32&25&18&15&12\\
\hline
$i$ &26&27&28&29&30\\
\hline
$t_i$ & 12&13&14&15&16\\
\hline
$x_{t_i}$&10&7&7&7&6\\
\hline
\end{tabular}
\end{table}
\textup{According to the equation (\ref{opt}), without loss of generality let $n=10$, the estimate $\tilde{\mu}_{0}$ solves the optimization problem
$$\min_{\mu_{0}}\sum_{i=1}^{9}\left(x_{t_{i+1}}-x_{t_i}-(k_{0}\mu_{0}-k_{1} x_{t_i})(t_{i+1}-t_i)\right)^2$$
which gives
$$\tilde{\mu}_{0}=\left(\sum_{i=1}^9(x_{t_{i+1}}-x_{t_{i}}+k_{1}x_{t_{i}}(t_{i+1}-t_{i}))\cdot(t_{i+1}-t_{i})\right)\times$$$$\left(\sum_{i=1}^{9}k_{0}(t_{i+1}-t_{i})^2\right)^{-1}$$
from the observational data
$$\tilde{\mu}_{0}=160.7381.$$
Then according to the equation (\ref{si}), the parameter $\tilde{\sigma}_{0}$ satisfies
\begin{align*}
&\sum_{i=1}^{9}\sigma_{0}^2\cdot(t_{i+1}-t_i)^2\\
=&\sum_{i=1}^{9}\left(x_{t_{i+1}}-x_{t_i}-(k_{0}\tilde{\mu}_{0}-k_{1}x_{t_{i}})(t_{i+1}-t_i)\right)^2
\end{align*}
that is,
$$\tilde{\sigma}_{0}^2=\left(\sum_{i=1}^{9}\left(x_{t_{i+1}}-x_{t_i}-(k_{0}\tilde{\mu}_{0}-k_{1}x_{t_{i}})(t_{i+1}-t_i)\right)^2\right)\times$$$$\left(\sum_{i=1}^{9}(t_{i+1}-t_i)^2\right)^{-1}$$
which gives $$\tilde{\sigma}_{0}=35.9460.$$
We can get the estimate values $(\tilde{\mu}_{0.25}, \tilde{\sigma}_{0.25})$, $(\tilde{\mu}_{0.30}, \tilde{\sigma}_{0.30})$, $\cdots$, $(\tilde{\mu}_{7}, \tilde{\sigma}_{7})$,
shown in Table \ref{T2}}.
\begin{table}
\caption{Estimated values for $\mu_{t_{m}}$ and $\sigma_{t_{m}}$ in Example \ref{ex3}}\label{T2}
\center
\begin{tabular}{|c|c|c|c|c|c|}
\hline
$m$ &1&2&3&4&5\\
\hline
$t_m$ & 0&0.25&0.30&0.35&0.40\\
\hline
$\tilde{\mu}_{t_m}$& 160.7381&129.2143&70.4127&34.1437&25.0147\\
\hline
$\tilde{\sigma}_{t_m}$&35.9460&61.4496&38.0270&22.1365&16.6541\\
\hline
$m$ &6&7&8&9&10\\
\hline
$t_m$&0.45&0.50&0.55&0.60&0.70\\
\hline
$\tilde{\mu}_{t_m}$&14.6400&11.0753&9.2512&9.4430&8.3679\\
\hline
$\tilde{\sigma}_{t_m}$&9.5265&8.6240&8.5882&7.9862&7.6481\\
\hline
$m$ &11&12&13&14&15\\
\hline
$t_m$ & 0.75&1&1.5&2&3\\
\hline
$\tilde{\mu}_{t_m}$&6.7920&5.1905&3.1973&1.7262&0.3452\\
\hline
$\tilde{\sigma}_{t_m}$&6.6699&4.8444&3.7815&3.4547&2.5632\\
\hline
$m$ &16&17&18&19&20\\
\hline
$t_m$ &3.5&4&4.5&5&6\\
\hline
$\tilde{\mu}_{t_m}$&0.6455&0.7714&0.1732&0.3810&0.2857\\
\hline
$\tilde{\sigma}_{t_m}$ &2.0422&1.8642&1.2220&1.2365&1.1547\\
\hline
$m$ &21&&&&\\
\hline
$t_m$ &7&&&&\\
\hline
$\tilde{\mu}_{t_m}$&0.0952&&&&\\
\hline
$\tilde{\sigma}_{t_m}$ &1.0499&&&&\\
\hline
\end{tabular}
\end{table}
\textup{Use MATLAB software to make $\tilde{\mu}_{t_{m}}$ scatter diagram of $t_{m}$, as shown in Figure 1.
\begin{figure}[h]
    \centering
    \includegraphics[width=8cm,height=5cm]{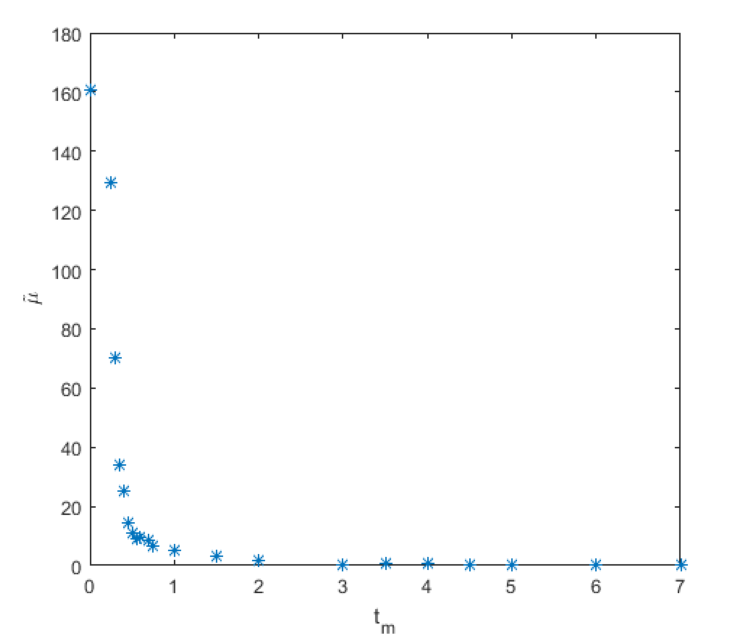}
    \caption{A scatter plot of $\tilde{\mu}$ with respect to $t_m$}
    \label{1}
\end{figure}
According to the trend of the scatter plot, we may employ Gaussian fitting function,
$$\mu_{t}=\hat{\beta}_{0}\exp(-((t-\hat{\beta}_{1})/\hat{\beta}_{2})^2)$$
where $\hat{\beta}_{0}$, $\hat{\beta}_{1}$, $\hat{\beta}_{2}$ are unknown parameters. The nonlinear fitting results are shown in Figure 2,
\begin{figure}[h]
    \centering
    \includegraphics[width=8cm,height=5cm]{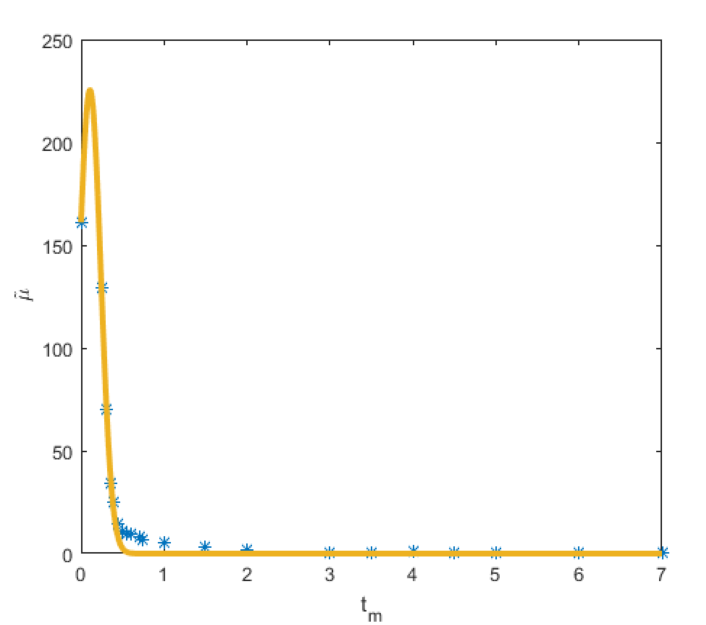}
    \caption{Gaussian fitting diagram}
    \label{2}
\end{figure}
where $R^2=0.9849$, by applying nonlinear least squares estimation, we get the time-varying parameters,
$\hat{\beta}_{0}=225.3$, $\hat{\beta}_{1}=0.1074$, $\hat{\beta}_{2}=0.1853$, then
$$\mu_{t}=225.3\exp(-((t-0.1074)/0.1853)^2).$$
And we get the $\sigma_{t}$,
$$\sigma_{t}=71.03\exp(-((t-0.1584)/0.1949)^2).$$
The uncertain differential equation,}
\begin{align*}
\intd X_t=(157.71\exp(-((t-0.1074)/0.1853)^2)-0.2X_t)\intd t\\+71.03\exp(-((t-0.1584)/0.1949)^2)\intd C_t .\\
\end{align*}
\end{exm}

\begin{exm}\label{ex4}
\textup{We known COVID-19 spread model based on uncertain differential equation,
$$\intd X_t=\mu_{t}X_t\intd t+\sigma_{t}X_t\intd C_t$$
where $\mu_{t}$ and $\sigma_{t}$ are  parameters to be estimated.
We used $N=35$ observations from COVID-19 shown in Table \ref{T3}. Let $t_{1}$, $t_{2}$, $\cdots$, $t_{35}$ represent the dates from February 13 to March 18, let
$x_{1}$, $x_{2}$, $\cdots$, $x_{35}$ represent the cumulative numbers on dates $t_{1}$, $t_{2}$, $\cdots$, $t_{35}$, respectively. }
\begin{table}
\caption{Observed data in Example \ref{ex4}}\label{T3}
\center\footnotesize
\begin{tabular}{|c|c|c|c|c|c|}
\hline
$i$ &1&2&3&4&5\\
\hline
$t_i$&1&2&3&4&5\\
\hline
$x_{t_i}$&63851&66492&68500&70548&72436\\
\hline
$i$ &6&7&8&9&10\\
\hline
$t_i$&6&7&8&9&10\\
\hline
$x_{t_i}$&74185&74576&75465&76288&76936\\
\hline
$i$ &11&12&13&14&15\\
\hline
$t_i$ &11&12&13&14&15\\
\hline
$x_{t_i}$&77150&77658&78064&78497&78824\\
\hline
$i$ &16&17&18&19&20\\
\hline
$t_i$ &16&17&18&19&20\\
\hline
$x_{t_i}$&79251&79824&80026&80151&80270\\
\hline
$i$ &21&22&23&24&25\\
\hline
$t_i$ &21&22&23&24&25\\
\hline
$x_{t_i}$&80389&80516&80591&80632&80668\\
\hline
$i$ &26&27&28&29&30\\
\hline
$t_i$ &26&27&28&29&30\\
\hline
$x_{t_i}$&80685&80699&80708&80725&80729\\
\hline
$i$ &31&32&33&34&35\\
\hline
$t_i$ &31&32&33&34&35\\
\hline
$x_{t_i}$&80733&80737&80738&80739&80739\\
\hline
\end{tabular}
\end{table}
\textup{According to the equation (\ref{opt}), the same $n=10$, the estimate $\tilde{\mu}_{1}$ solves the optimization problem
$$\min_{\mu_{1}}\sum_{i=1}^{9}\left(x_{t_{i+1}}-x_{t_i}-\mu_{1} x_{t_i}\cdot(t_{i+1}-t_i)\right)^2$$
according to the example 1, we get that the estimate of $\mu_{1}$ is
$$\left(\sum_{i=1}^{9}(x_{t_{i+1}}-x_{t_{i}})(t_{i+1}-t_{i})x_{t_{i}}\right)\cdot\left(\sum_{i=1}^{9}x_{t_{i}}^2(t_{i+1}-t_{i})^2\right)^{-1}$$
which gives
$$\tilde{\mu}_{1}=0.0198.$$ Then according to the example 1, the parameter $\tilde{\sigma}_{1}$ satisfies
\begin{align*}
&\sum_{i=1}^{9}\sigma_{1}^2\cdot x_{t_{i}}^2(t_{i+1}-t_i)^2\\
=&\sum_{i=1}^{14}\left(x_{t_{i+1}}-x_{t_i}-\tilde{\mu}_{1} x_{t_i}\cdot(t_{i+1}-t_i)\right)^2
\end{align*}
which gives $$\tilde{\sigma}_{1}=0.0113.$$
We can also get the estimate values $(\tilde{\mu}_{2},\tilde{\sigma}_{2})$, $(\tilde{\mu}_{3},\tilde{\sigma}_{3})$, $\cdots$, $(\tilde{\mu}_{26}, \tilde{\sigma}_{26})$,
shown in Table \ref{T4}. Use software to plot a scatter plot of $\tilde{\mu}_{t_{m}}$ with respect to $t_{m}$, as shown in Figure 3.
\begin{figure}[h]
    \centering
    \includegraphics[width=8cm,height=5cm]{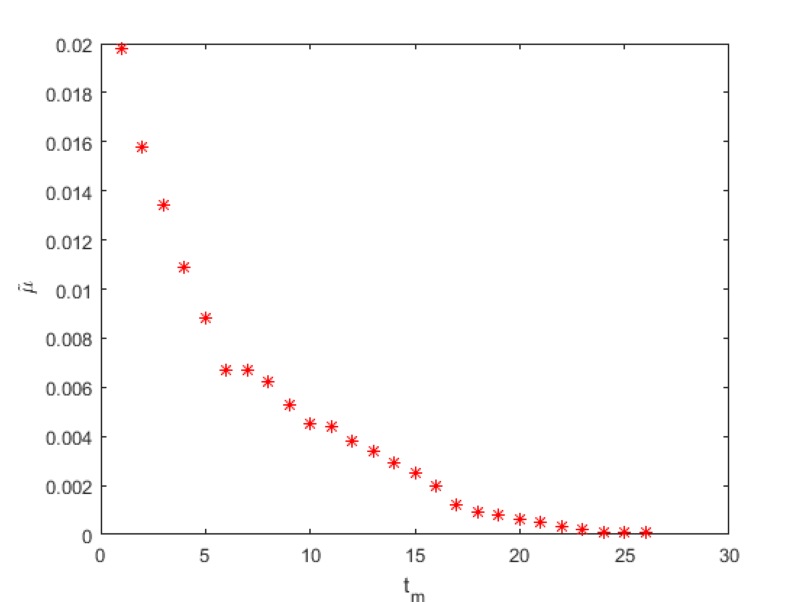}
    \caption{A scatter plot of $\tilde{\mu}$ with respect to $t_m$}
    \label{3}
\end{figure}
\\Next, the fitting regression was carried out for $\tilde{\mu}_{t_m}$ $(m=1, 2, \cdots, 26)$ in Table \ref{T4}.
\begin{table}
\caption{Estimated values for $\mu_{t_{m}}$ and $\sigma_{t_{m}}$ in Example \ref{ex4}}\label{T4}
\center
\begin{tabular}{|c|c|c|c|c|c|}
\hline
$m$ &1&2&3&4&5\\
\hline
$t_m$ & 1&2&3&4&5\\
\hline
$\tilde{\mu}_{t_m}$& 0.0198&0.0158&0.0134&0.0109&0.0088\\
\hline
$\tilde{\sigma}_{t_m}$&0.0113&0.0102&0.0093&0.0078&0.0058\\
\hline
$m$ &6&7&8&9&10\\
\hline
$t_m$&6&7&8&9&10\\
\hline
$\tilde{\mu}_{t_m}$&0.0067&0.0067&0.0062&0.0053&0.0045\\
\hline
$\tilde{\sigma}_{t_m}$&0.0029&0.0028&0.0022&0.0018&0.0018\\
\hline
$m$ &11&12&13&14&15\\
\hline
$t_m$ & 11&12&13&14&15\\
\hline
$\tilde{\mu}_{t_m}$&0.0044&0.0038&0.0034&0.0029&0.0025\\
\hline
$\tilde{\sigma}_{t_m}$&0.0020&0.0020&0.0021&0.0021&0.0021\\
\hline
$m$ &16&17&18&19&20\\
\hline
$t_m$ &16&17&18&19&20\\
\hline
$\tilde{\mu}_{t_m}$&0.0020&0.0012&0.0009&0.0008&0.0006\\
\hline
$\tilde{\sigma}_{t_m}$ &0.0019&0.0007&0.0006&0.0006&0.0005\\
\hline
$m$ &21&22&23&24&25\\
\hline
$t_m$ &21&22&23&24&25\\
\hline
$\tilde{\mu}_{t_m}$&0.0005&0.0003&0.0002&0.0001&0.0001\\
\hline
$\tilde{\sigma}_{t_m}$ &0.0005&0.0003&0.0002&0.0001&0.0001\\
\hline
$m$ &26&&&&\\
\hline
$t_m$ &26&&&&\\
\hline
$\tilde{\mu}_{t_m}$&0.0001&&&&\\
\hline
$\tilde{\sigma}_{t_m}$ &0.0001&&&&\\
\hline
\end{tabular}
\end{table}
According to the trend of the scatter plot, we may employ logistic decay model,
$$\mu_{t}=\frac{0.0198}{1+\hat{\beta}_{0}\exp(\hat{\beta}_{1}t)}$$
where $\hat{\beta}_{0}$ and $\hat{\beta}_{1}$ are unknown parameters, the nonlinear fitting results are shown in Figure 4,
\begin{figure}[h]
    \centering
    \includegraphics[width=8cm,height=5cm]{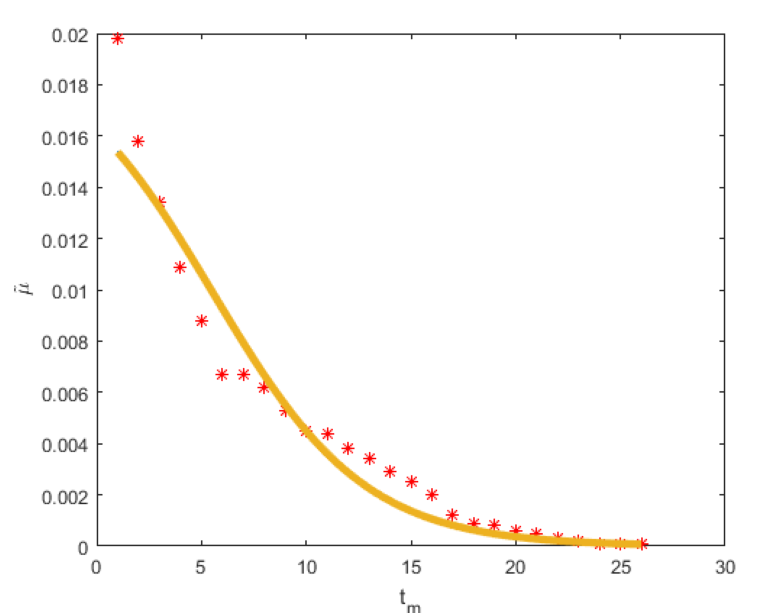}
    \caption{Logical decline model fitting}
    \label{4}
\end{figure}
we get the time-varying parameters, $\hat{\beta}_{0}=0.2190$, $\hat{\beta}_{1}=0.2745$, then
$$\mu_{t}=\frac{0.0198}{1+0.2190\exp(0.2745t)}$$
And we get the $\sigma_t$,
$$\sigma_{t}=\frac{0.0113}{1+0.0894\exp(0.4471t)}$$
therefore, the COVID-19 spread model based on uncertain differential equation,
$$\intd X_t=\frac{0.0198X_t\intd t}{1+0.2190\exp(0.2745t)}+\frac{0.0113X_t\intd C_t}{1+0.0894\exp(0.4471t)}.$$}
\end{exm}

\section{Conclusions}
Parameter estimation of uncertain differential equations is a very important problem. The method of the least squares estimation was employed in this paper to estimate the time-varying parameters in uncertain differential equation. Based on the least square estimation method, the paper first obtains the value of the parameters at a fixed time, then obtains a set of parameter estimates as time goes on, and then gets the estimation equation of the time-varying parameters by regression analysis fitting. Using this method, the propagation model of COVID-19 based on uncertain differential equations is obtained. There remain many research problems in this area, for example, how to estimate the time-varying parameters in uncertain differential equations by means of maximum likelihood estimation, by means of generalized moment estimation, by means of the discretely sampled data via the $\alpha$-path.

\section{Acknowledgements}
This research was funded by the Natural Science Foundation of Xinjiang (Grant No. 2020D01C017) and the National Natural Science Foundation of China (Grant Nos. 12061072 and U1703262).

\ifCLASSOPTIONcaptionsoff
  \newpage
\fi

\begin{IEEEbiography}
[{\includegraphics[width=1.2in,height=1.35in,clip,keepaspectratio]{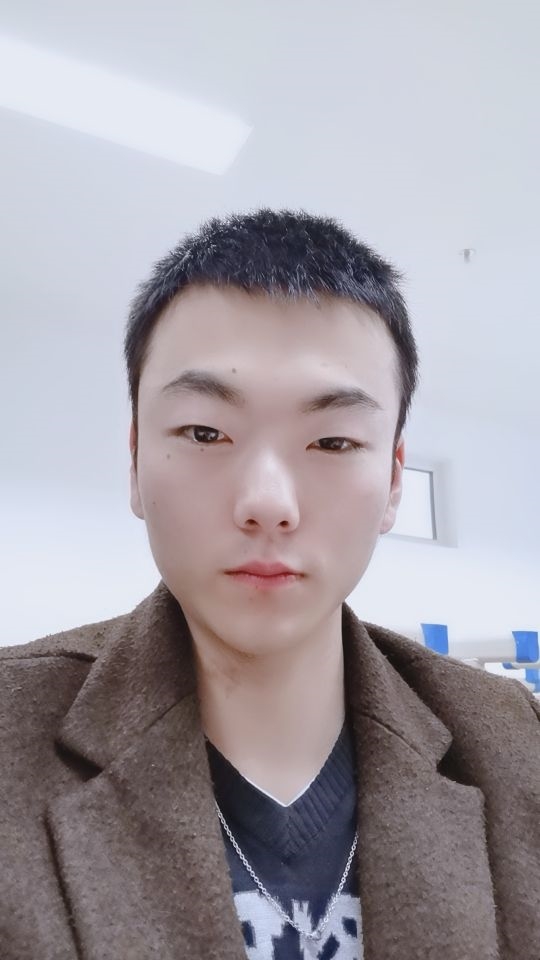}}]{Guidong Zhang}
received the B.S. degree form Yili Normal University, Yining, China, in 2016. He is currently a student at the College of Mathematical and System Sciences, Xinjiang University, Urumqi 830046, China. His current research interest is uncertain differential equations.
\end{IEEEbiography}

\begin{IEEEbiography}
[{\includegraphics[width=1.2in,height=1.35in,clip,keepaspectratio]{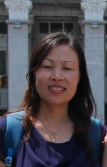}}]{Yuhong Sheng}
received the Ph.D. degree from Tsinghua University, Beijing, China, in 2015. She is currently an Associate Professor with the College of Mathematical and System Sciences, Xinjiang University, Urumqi 830046, China. She has authored or coauthored about 30 articles on several journals including Applied Mathematics and Computation, Fuzzy Optimization and Decision Making, Computers and Industrial Engineering, Soft Computing, Journal of Intelligent \& Fuzzy Systems, Journal of Uncertainty Analysis and Applications, Chaos Solitons \& Fractals, IEEE Transactions on Fuzzy Systems, Mathematical Methods in the Applied Sciences, and Journal of Uncertain Systems. Her current research interests include uncertain process, uncertain systems, uncertain differential equations, chance theory and their applications.
\end{IEEEbiography}


\begin{thebibliography}{1}
\bibitem{Liu1} Liu, B. Uncertainty Theory, second, Spring-Verlag, Berlin, 2007.

\bibitem{Liu2} Liu, B. Fuzzy process, hybrid process and uncertain process. Uncertain Syst. 2(1) (2008) 3-16.

\bibitem{Liu3} Liu, B. Toward uncertain finance theory. J. Uncertain. Anal. Appl. 1 (2013) 1.
\bibitem{Liu4} Liu, B. Some research problems in uncertainty theory. J. Uncertain  Syst. 3(1) (2009) 3-10.

\bibitem{Zhu} Zhu, Y. Uncertain optimal control with application to a portfolio selection model. Cybern Syst. 41(7) (2010) 535-547.

\bibitem{Yang1} Yang, X., Yao, K. Uncertain partial differential equation with application to heat conduction. Fuzzy Optim Decis Making. 16(3) (2017) 379-403.

\bibitem{Zhang}  Zhang, Z., Yang, X. Uncertain population model. Soft Comput. 24(4) (2020) 2417-2423.

\bibitem{Yao1} Yao, K. Uncertain differential equations, New York, Springer, 2016.

\bibitem{Sheng1} Sheng, Y., Yao, K., Chen, X.  Least squares estimation in uncertain differential equations. IEEE Trans Fuzzy Syst. 28(10) (2020) 2651-2655.

\bibitem{Yao2} Yao, K., Liu, B. Parameter estimation in uncertain differential equations. Fuzzy Optim Decis Making. 19(1) (2020) 1-12.

\bibitem{Liu5} Liu, Z.  Generalized moment estimation for uncertain differential equations. Appl Math Comput. 392 (2021) 125724.

\bibitem{Lio1} Lio, W., Liu, B. Initial value estimation of uncertain differential equations and zero-day of COVID-19 spread in China. Fuzzy Optim Decis Making. (2020).

\bibitem{Lio2} Lio, W., Liu, B. Uncertain maximum likelihood estimation with application to uncertain regression analysis. Soft Comput. 24 (2020) 9351-9360.

\bibitem{Yang2} Yang, X., Liu, Y.,  Park, G.K.  Parameter estimation of uncertain differential equation with application to financial market. Chaos Solitons Fractals. 139 (2020) 110026.

\bibitem{Sheng2} Sheng, Y., Zhang, N. Parameter estimation in uncertain differential equations based on the solution. Math Meth Appl Sci. (2021) 1-12.


\bibitem{Zhang1} Zhang, J., Sheng, Y. Least squares estimation of uncertain delay differential equations. (2021) Technical Report.


\bibitem{Zhang2} Zhang, J., Sheng, Y., Wang, X. Least squares estimation of high-order uncertain differential equations. (2020) Technical Report.


\bibitem{Sheng4} Sheng, Y., Wang, C. Stability in the p-th moment for
 uncertain differential equation. J Intell Fuzzy Syst. 26(3) (2014)
1263-1271.
 \bibitem{Sheng5} Sheng, Y. Stability of high-order uncertain
differential equations. J Intell Fuzzy Syst. 33 (2017) 1363-1373.

\bibitem{Liu6} Liu, Z., Yang, Y.  Moment estimation for parameters
in high-order uncertain differential equations. Appl Math
Comput. 392 (2019) 125724.



\bibitem{Yao3} Yao, K., Chen, X., A numerical method for solving uncertain differential equations. J Intell Fuzzy Syst. 25(3) (2013) 825-832.

\bibitem{Yang3} Yang, X., Gao, J. Uncertain differential games with application to capitalism. J. Uncertain. Anal. Appl. 1 (2013) 17.

\bibitem{Chen1} Chen, X.  American option pricing formula for uncertain financial market, Int J Oper Res. 8(2) (2011) 27-32.

\bibitem{Chen2}Chen, X., Liu, B.  Existence and uniqueness theorem for uncertain differential equations, Fuzzy Optim Decis Making. 9(1) (2010) 69-81.
\bibitem{Yao4}Yao, K., Gao, J., Gao, Y.  Some stability theorems of uncertain differential equation. Fuzzy Optim Decis Making. 12(1) (2013) 3-13.

\bibitem{Liu7}Liu, Y. Analytic method for solving uncertain differential equations. J. Uncertain. Syst. 6 (2012) 243-248.
\bibitem{Zhang3} Zhang, Z., Liu, W. Geometric average asian option pricing for uncertain financial market. J. Uncertain Syst. 8(4) (2014) 317-20.
\end{thebibliography}
\end{document}